\documentstyle[11pt,paspconf]{article}

\begin{document}

\title{Evolution of Gas and Dust in Circumstellar Disks}

\author{David W. Koerner}
\affil{Dept.\ of Physics and Astronomy, University of Pennsylvania,
Philadelphia, PA 19104-6396}



\begin{abstract}

A clear understanding of the chemical processing of matter as it is
transferred from a molecular cloud to a planetary system depends heavily 
on the physical conditions endured by gas and dust as these accrete
onto a disk and are incorporated into planetary bodies. Reviewed here are 
astrophysical observations of circumstellar disks which trace their
evolving properties. Accretion disks that are
massive enough to produce a solar system like our own
are typically larger than 100 AU. This suggests that the chemistry of 
a large fraction of the infalling material is not
radically altered upon contact with a vigorous accretion shock. 
The mechanisms of accretion onto the star
and eventual dispersal are not yet well understood,
but timescales for the removal of gas and optically
thick dust appear to be a few times 10$^6$ yrs. At later times, 
tenuous ``debris disks'' of dust remain around stars as old as a 
few times 10$^8$ yrs. Features in the morphology of the
latter, such as inner holes, warps, and azimuthal asymmetries, 
are likely to be the result of
the dynamical influence of large planetary bodies. 
Future observations will enlighten our understanding of chemical
evolution and will focus on the search
for disks in transition from a viscous accretion stage to 
one represented by a gas-free assemblage of colliding planetesimals. 
In the near future, 
comparative analysis of circumstellar dust and gas properties 
within a statistically significant sample of young stars at various ages
will be possible with instrumentation such as SIRTF and SOFIA. 
Well-designed surveys will
help place solar system analogs in a general context of 
a diversity of possible pathways for circumstellar evolution, one
which encompasses the formation of stellar and brown-dwarf companions
as well as planetary systems.

\end{abstract}



\section{Introduction}

The chemical history of solar system materials took place within
a wide diversity of physical and chemical environments, as
gas and dust were transported from a molecular cloud core to a destination 
in planetary bodies and atmospheres.  Until recently, reconstruction 
of the relevant initial conditions and evolutionary processes 
relied solely on vestigial evidence wrested from analyses
of the chemical composition of matter in the present-day solar system.
This method of investigation is limited, however, by difficulties inherent in
bridging a temporal discontinuity of over 4 billion years. 
It is especially challenging to disentangle the signatures of antecedent 
conditions from those of more recent processes occurring 
within the solar system (E.g., Brown, this volume). 
Improvements in observational methods have now made
it possible to leap across this gap in time and
directly observe analogs of the milieu in 
which early solar system materials were forged.
Circumstellar disks and envelopes around currently-forming young stars
have become the subject of increasingly detailed
investigations with high-resolution astronomical 
techniques operating across a broad range of wavelengths 
(See reviews by Beckwith \& Sargent 1996 and Koerner 1997).

The presence of protostellar envelopes and accretion disks around
young stars was disclosed as soon as appropriate
long-wavelength detectors became available. Observations extending 
from near infrared to millimeter 
wavelengths revealed excess radiation with a range of 
properties that implied an
evolutionary sequence of states for the circumstellar material
(Myers \& Benson 1983; Lada \& Wilking 1984;
Strom et al.\ 1989; Beckwith et al.\ 1990).
Models of the associated ``Spectral Energy Distributions'' (SEDs)
accounted for diverse spectral shapes by locating dust at
various distances from the young star, partitioned between
spherical, flared disk, or flat disk configurations 
(Adams, Lada, \& Shu 1987; Kenyon \& Hartmann 1987; 
cf.\ Beckwith \& Sargent 1993 and Shu et al.\ 1993).

The first images to confirm the above interpretations
also validated the idea that differing SED properties were
due to the time dependence of circumstellar disk properties.
Aperture synthesis imaging
of CO(2$\rightarrow$1) emission from the very young ($t \sim 10^5$ yr)
stellar object, HL Tauri, revealed circumstellar 
dust and gas in an elongated structure with a mass several times that
of the minimum required to form a solar system like our own (Beckwith et al.\
1986; Sargent \& Beckwith 1987; 1991).  Coronagraphic imaging
of dust around the much older  ($t \sim 10^{7-8}$~yr)
main sequence star, $\beta$ Pictoris,
revealed a far more tenuous dust disk 
(Smith \& Terrile 1984). The small
mass of material and its short lifetime against dispersal
implied that any associated planet formation had largely taken
place (cf.\ Backman \& Paresce 1993). It now appears that these 
objects represent snapshots at times which bracket most of the
early evolution of protoplanetary disk systems. Gaps in the 
sequence are rapidly being filled in with new high-quality images
of a wide range of objects.

\section{Imaging the Stages of Disk Evolution}

\subsection{The Embedded Protostar Stage}
 
\begin{figure}
\plotfiddle{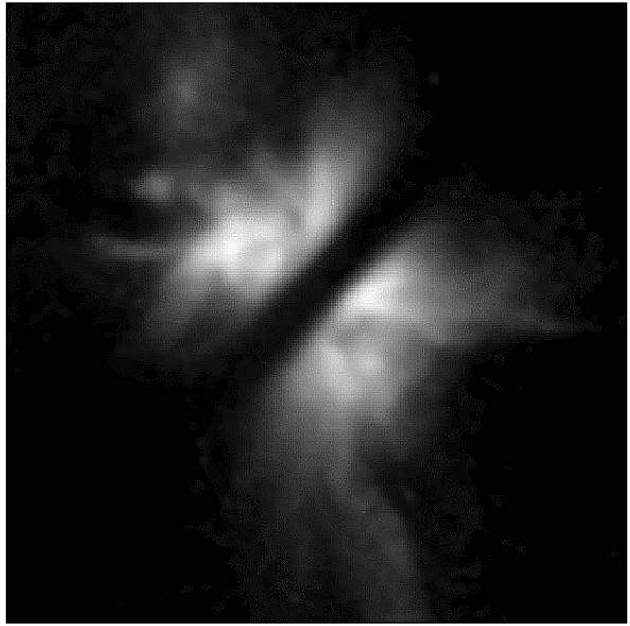}{220pt}{0}{40}{40}{-130}{-10}
\caption {HST/NICMOS image of the nebulosity in IRAS 04302 (taken from
Padgett et al.\ 1999). The dust lane, corresponding to an edge-on
disk embedded within a flattened envelope, is nearly 1000 AU in diameter.}
\end{figure}

Chemical processing of protostellar and protoplanetary
gas and dust begins even before it encounters a
shock front at the disk-envelope interface. 
Radiation from the central protostar, enhanced 
density in the molecular cloud core, and interaction with
far-reaching ionized jets and bipolar outflows all contribute to
distort the original interstellar chemical signature
(Bergin, H\"ogerheijde, and Ohashi, this volume). These changes may
vary systematically in a time-dependent way, making it possible
to use chemical abundances as a chronometer which traces 
the evolution of the infalling envelope (cf. Langer et al.\ 2000). 

As infalling matter impacts a circumstellar accretion disk, it 
is subject to shock heating and gas drag with an intensity that depends 
sensitively on the radial distance of the impact from
the star and concomitant impact velocity (see review by Lunine 1997 and
references therein).
In particular, icy grains accreting at distances greater than 30 AU from
a Sun-like star are likely to suffer little sublimation of volatiles, and 
gas molecules are unlikely to be dissociated. 
Consequently, disk material originally at this radius is unlikely to 
bear much of the imprint of its entry into the disk.

High-resolution images now reveal that much of the material incorporated
into circumstellar disks does indeed originally arrive at radial distances
much greater than 30 AU. Although much of the material in the outer regions 
of the flattened structure around HL Tauri is now known to be infalling 
(Hayashi et al.\ 1993), continuum images indicate the presence of
a central disk, presumably
centrifugally supported, with a radius of order 100 AU (Lay et al.\ 1994;
Mundy et al.\ 1996). Flattened structures of similar size and 
kinematics have recently been imaged around a small sample of other embedded
young stars, in both CO line emission (cf.\ Ohashi, this volume), and 
in scattered light (Padgett et al.\ 1999). One example, IRAS 04302,
is displayed in Fig.\ 1 and appears as a 450-AU-radius circumstellar 
structure oriented edge on with a highly flattened 
morphology (Padgett et al.\ 1999). Kinematic
analysis of CO spectral line images indicates that 
rotational motions dominate the velocity field of this structure. 
These results strongly suggest that
most of the gas in solar nebula analogs originally arrives at distances 
greater than
100 AU where little processing takes place. The large sizes of more-evolved
centrifuglly supported disks bear this out.

\subsection{ The ``T-Tauri'' Phase}

Young stars first become optically visible when their infall envelope 
has dispersed enough to become transparent.
Surveys of unresolved infrared and millimeter-wave 
emission from these ``T Tauri stars'' (TTs) provided initial evidence
that circumstellar disks are the dominant component in
the dust configuration at this stage
(Strom et al.\ 1989; Beckwith et al.\ 1990). A large fraction of
TTs are detected with associated SEDs
that can be attributed to dusty disks, similar to those expected for 
planetary systems in formation (cf.\ Beckwith \& Sargent 1993).
The overwhelming majority of these are associated with ``classical''
T Tauri stars (cTTs), for which diagnostics of protostellar accretion such as
H$\alpha$ emission are still robust (cf. Calvet, Hartmann, \& Strom 2000).
Optically thick dust emission is not as readily detected
from disks around T Tauri stars without strong evidence of 
protostellar accretion, the so-called
``weakline T Tauri stars'' (wTTs) (cf.\ Osterloh \& Beckwith 1995).  

\begin{figure}
\plotfiddle{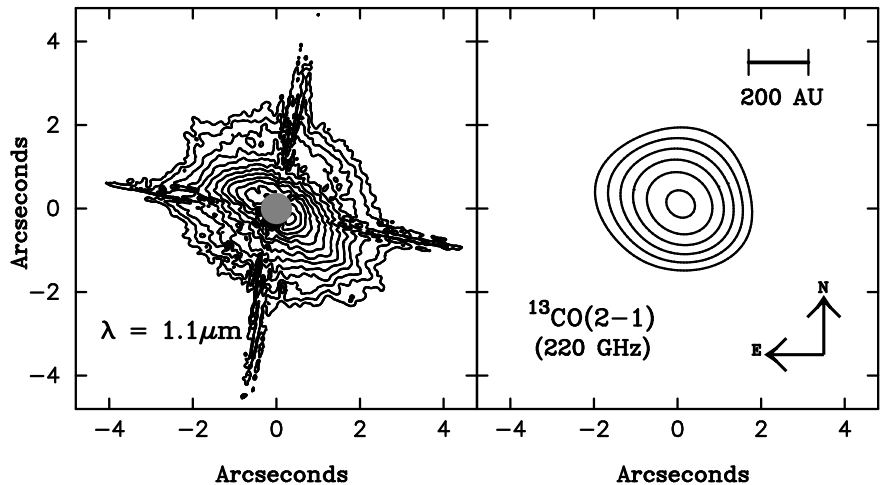}{200pt}{270}{50}{50}{-200}{250}
\caption {(Left) HST/NICMOS corongraphic image of the cTTs GM Aurigae and 
(right) OVRO aperture synthesis image of $^{13}$CO(2$\rightarrow$1)from the
same star. The HST image reveals a flattened circumstellar reflection 
nebula extending symmetrically from the star to radial distances
of 3$''$ (450 AU) and aligned with the long axis of molecular emission.
Kinematic analysis of the gas establishes that the disk is in Keplerian
rotation about the star.}
\end{figure}

Aperture synthesis imaging of the disk around GM Aurigae was the
first to demonstrate that gas was rotationally supported throughout
the radial extent of a circumstellar disk around a cTTs
(Koerner et al.\ 1993). High-resolution coronagraphic imaging
with HST confirms the picture from mm-wave interferomery and is displayed
in Fig.\ 2 (Koerner et al.\ 1999). Light scattered from the concave surface
of a flared disk is consistent with an orientation like that derived from
aperture synthesis images of the molecular emission. 
The disk is several hundred AU in radius and has a mass several times that 
required to form our own solar system (cf.\ Dutrey et al.\ 1998).

Additional observations of CO emission from cTTs have demonstrated that
the properties of the disk around GM Aur are not at all unusual. 
TTs which exhibit similar mm-wave continuum luminosity are typically 
surrounded by disks with radii greater than 100 AU (Koerner
\& Sargent 1995; Dutrey, this volume). The occurrence frequency for
mm-wave detection at the associated luminosity level
is about 10$\%$ for TTs, generally. These are the only objects 
which can be appropriately considered to be analogs
of the early solar nebula, since disk 
masses derived for TTs with lower mm-wave continuum luminosity are well 
below the minimum required to produce a solar system like our own.

The timescale and associated mechanism by which disks eventually 
disperse is not conclusively determined. Photo-evaporation may provide
a way to deplete disk gas from the outside in (see Johnstone, this volume),
while accretion onto the star may remove material from the inner
disk.  Infrared surveys establish that inner-disk dust is 
depleted in TTs older than 3 $\times 10^6$ yrs 
(Skrutskie et al.\ 1990), but it is unclear whether the 
underlying cause is pre-planetary grain accumulation, protostellar accretion, 
or some other dispersal mechanism.  Surveys at millimeter and 
sub-millimeter wavelengths fail to reveal a correlation between
stellar age and associated disk mass (Beckwith et al.\ 1990; 
Beckwith \& Sargent 1991), but such an effect may be masked by the failure of 
studies to discriminate between single and binary stars
(Jensen et al.\ 1994; Osterloh \& Beckwith 1995).
Evidence for the persistence time of gas in disks is scarcer than for dust,
even though 99\% of the mass of protoplanetary disks is thought to 
consist of molecular gas. There is some suggestion that gaseous disks are 
dispersed on timescales of 10$^7$--10$^8$ years (Skrutskie et al.\ 1991;
Zuckerman, Forveille, \& Kastner 1995), but the number of objects 
observed with sufficient sensitivity is still quite small. In any event, 
gaseous disks similar to those imaged around cTTs have not been 
imaged for any weakline T Tauri stars, except for couple of 
borderline cases (Duvert et al.\ 1999; Qi, this volume).

\subsection{Debris Disks}

\begin{figure}
\plotfiddle{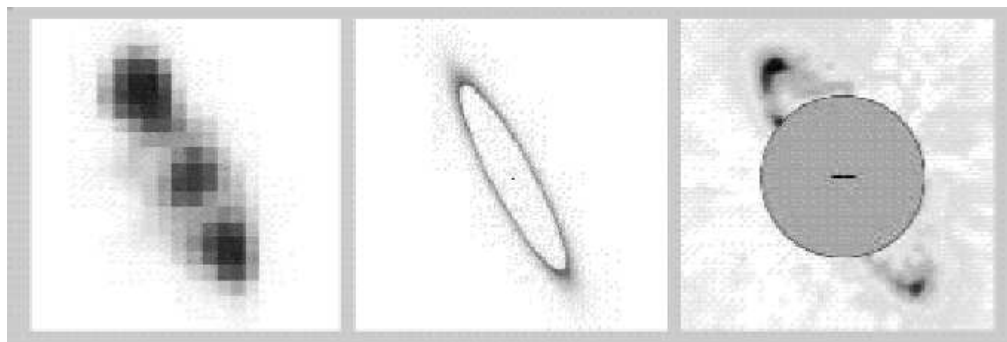}{120pt}{0}{65}{65}{-200}{-175}
\caption {(Left) Keck/MIRLIN image of HR 4796A at 24.5 $\mu$m. The elongated
structure is $\sim$2$''$ in diameter, 
corresponding to $\sim$150~AU. (Center) A model
of the underlying emission structure that was obtained by fitting to
an image at 20 $\mu$m from Koerner et al.\ (1998). 
(Right) HST/NICMOS coronagraphic image of 
scattered light from the ring
around HR 4796A at $\lambda$ = 1.1 $\mu$m taken from Schneider et al.\ (1999).
}
\end{figure}

The presence of remnant circumstellar dust around ``Vega-type'' 
stars --  A-type stars with infrared excess and
ages between ten and a few hundred 
million yrs -- may signal a more evolved stage of planet formation
(cf.\ Backman \& Paresce 1993; Lagrange, Backman, \& Artymowicz 2000
and references therein). The infrared signature of optically thin
dust is present in IRAS measurements, but the correlation with
early spectral type may be due simply to the ability of hotter stars
to heat circumstellar dust at several 10's of AU to 
temperatures characteristic of infrared radiation.
Images of these tenuous ``debris disks'' have been obtained in both 
scattered light and thermal infrared emission for 
$\beta$ Pictoris (Smith \& Terrile 1984; Lagage \& Pantin 1994;
Heap et al.\ 1997), HR 4796A (Jayawardhana et al.\ 1998; 
Koerner et al.\ 1998; Schneider et al.\ 1999), HD 141569
(Augereau et al.\ 1999; Weinberger et al.\ 1999), and at sub-millimeter 
wavelengths for several nearby stars (Holland et al.\ 1998; 
Greaves et al.\ 1998). In many cases, the images confirm what was deduced 
by models of the spectral distribution of radiated energy, namely that
the disks surround large inner holes with sizes like that of our
solar system. This is readily apparent in thermal IR and HST images
of HR 4796A shown in Figure 3, where the dust is confined largely to a 
circumstellar ring.

The presence of a large hole in the disk around HR 4796 was originally
implied by the shape of its SED (Jura et al.\ 1995), a characteristic
that applies to many other debris-disk examples as well.
Modeling of thermal infrared imaging like that shown in Fig.\ 3
demonstrated unequivocally that the disk did not extend all the
way to the star (Koerner et al.\ 1998). This result was dramatically
confirmed in coronagraphic imaging with the Hubble Space Telescope
(Schneider et al.\ 1999), also displayed in Fig.\ 3. New analysis of 
imaging at 24.5 $\mu$m reveals that most of the emission at the 
stellar position is well in excess of the photosphere. The color
temperature of dust close to the star is $\sim$ 170 K, similar to
that expected for an ice condensation front like that which may
have assisted in the formation of Jupiter. For HR 4796A, a star
considerably warmer than the Sun, this temperature corresponds to
a a radial distance of about 10 AU (Wahhaj et al.\ 2000).

\begin{figure}
\plotfiddle{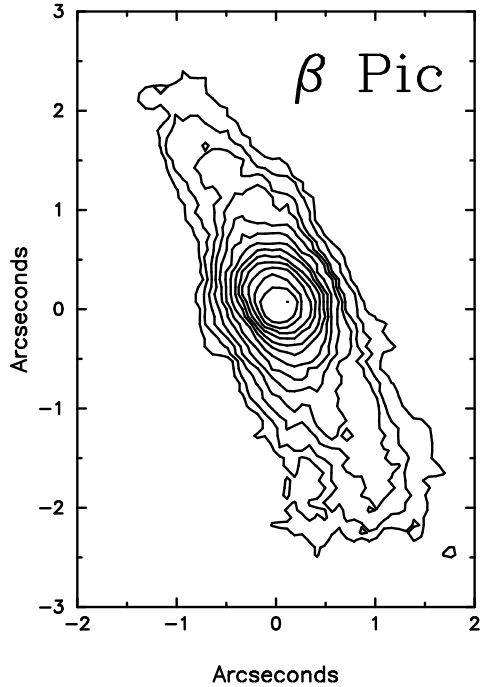}{210pt}{270}{50}{50}{-200}{300}
\caption {Keck/MIRLIN image of $\beta$ Pictoris at $\lambda$ = 20 $\mu$m.
The long axis of intermediate contours is slightly offset relative to
other contours, in keeping with the discovery of a warp in the inner disk.
This warp may signal the presence of a planetary body with an orbit
that is inclined relative to the circumstellar disk plane.}
\end{figure}

The presence of an inner hole in the disk around $\beta$ Pictoris is
also implied by its SED (Beckman et al.\ 1992), but is not quite as
readily apparent as for HR 4796A. This is clear from the
thermal infrared image in Fig.\ 4. The appearance of continuity
is deceiving, however,
since the emission intensity depends heavily on temperature and
will be preferentially greater for the material close to the star. 
Modeling of such images indicates that an inner region of reduced
density is indeed present (cf.\ Lagage \& Pantin 1994; Pantin et al.\ 1997).
Additional evidence for planetary bodies is apparent in the form of
a warp sharply identified in HST images (Burrows et al.\ 1995; 
Heap et al.\ 1997).
It appears in Fig.\ 4 as a variation in the angle of the
long axis of intermediate contour levels. The warp may be the result
of the dynamic influence of a planetary body with an orbit which is
inclined with respect to the plane of the disk.

Holes and/or gaps are evident in a few other extant disk images,
including those for $\alpha$ PsA, $\epsilon$ Eri, and HD 141569
(Holland et al.\ 1998; Greaves et al.\ 1998; Augereau et al.\ 1999;
Weinberger et al.\ 1999). These features strengthen the interpretation
of debris disks as representing a late protoplanetary phase in which
the system is largely devoid of molecular gas and contains 
fully formed planets and/or planetesimals which generate 
remnant debris via mutual collisions. 
The implied connection between disks and planets
has become even more explicit with the ground-based coronagraphic
detection of dust around a star for which a planet has actually 
been detected. Dust detected around 55 Cnc, a star with a radial-velocity
signature of a planet (Marcy et al.\ 1999), lies well outside of the orbit of
the detected body, but implies an orbital plane which confirms a
planetary mass for the companion (Trilling \& Brown 1998).

\section{Discussion}

The images reviewed above help establish and refine 
a picture of the evolution of
protoplanetary disks that is inferred from long-wavelength spectral
properties. It is now clear that the typical size for solar-system-analog
disks is larger than the canonical solar system (R $\sim$ 50 AU) by
factors of several. This suggests that a large fraction of the initial
molecular reservoir was not heavily modified by an accretion shock. 
The timescales for survival of optically thick dust and molecular
gas seems to be similar, of order a few times 10$^6$ yr. Disks of
tenuous debris frequently survive for another 100 million years and
show evidence of the dynamic influence of larger bodies.

Many questions remain unanswered in the above picture. 
It would be especially
useful to obtain images of disks in transition between a
viscously accreting stage and one in which gas and dust 
are largely dispersed. These would help refine estimates of the timescales
involved and could enlighten our understanding of the dispersal 
mechanisms as well. In addition, it is not always clear whether 
differences between individual disks are the result of evolution or
simply of different initial conditions. In order 
to sort this out, the statistical properties
of circumstellar matter around a large unbiased sample of young stars
must be obtained. High-resolution 
imaging of such a sample is currently infeasible,
but broadband spectral characteristics will be accessible to infrared
surveys taken with instruments such as SIRTF and SOFIA. These are designed
to operate above the atmosphere with the sensitivity required for detection
of waning disks in star-forming regions. Interpretations of such
surveys will,
of course, be helped by the ``ground truth'' afforded by available
images, but they may not require detailed imaging of every source. It
is expected that the broadband spectral properties of 
hundreds of young stars can be surveyed
with currently planned instrumentation. This will allow us to begin to 
answer
questions, not just about the evolution of solar system analogs, but
about the place of our solar system within a 
diversity of possible circumstellar environments.

%
%

%

\end{document}